\documentclass[apj]{emulateapj}
\usepackage{times}
\usepackage{amsfonts}
\usepackage{graphicx,times}
\usepackage[colorlinks=true,linkcolor=black,citecolor=black,urlcolor=blue]{hyperref}

\newcommand{\msunh}{\>h^{-1}\rm M_\odot}

\newcommand{\mpch}{\>h^{-1}{\rm {Mpc}}}
\newcommand{\kms}{\>{\rm km}\,{\rm s}^{-1}}
\newcommand{\kmsmpc}{\>{\rm km}\,{\rm s}^{-1}\,{\rm Mpc}^{-1}}

\newcommand{\rmag}{\>^{0.1}{\rm M}_r-5\log h}

\usepackage{xcolor}

\shorttitle{GALAXY GROUPS: THE 2MRS CATALOG}

\shortauthors{Lu et al.}

\begin{document}


\title{GALAXY GROUPS IN THE 2MASS Redshift Survey}

\author{Yi Lu\altaffilmark{1}, Xiaohu Yang\altaffilmark{2,3}, Feng
  Shi\altaffilmark{1}, H.J. Mo\altaffilmark{4,5}, Dylan
  Tweed\altaffilmark{2}, Huiyuan Wang\altaffilmark{6}, Youcai
  Zhang\altaffilmark{1}, Shijie Li\altaffilmark{1}, S.H.
  Lim\altaffilmark{4}}

\altaffiltext{1}{Key Laboratory for Research in Galaxies and
  Cosmology, Shanghai Astronomical Observatory, Nandan Road 80,
  Shanghai 200030, China; E-mail:
  \href{malto:luyi@shao.ac.cn}{luyi@shao.ac.cn}}

\altaffiltext{2}{Center for Astronomy and Astrophysics, Shanghai Jiao
  Tong University, Shanghai 200240, China; E-mail:
  \href{malto:xyang@sjtu.edu.cn}{xyang@sjtu.edu.cn}}

\altaffiltext{3}{IFSA Collaborative Innovation Center, Shanghai Jiao
  Tong University, Shanghai 200240, China}

\altaffiltext{4} {Department of Astronomy, University of
  Massachusetts, Amherst MA 01003-9305, USA}

\altaffiltext{5} {Physics Department and Center for Astrophysics, Tsinghua University,
  Beijing 10084, China}

\altaffiltext{6}{Key Laboratory for Research in Galaxies and
  Cosmology, Department of Astronomy, University of Science and
  Technology of China, Hefei, Anhui 230026, China}


\begin{abstract}
  A galaxy group catalog is constructed from the 2MASS Redshift Survey
  (2MRS) with the use of a halo-based group finder.  The halo mass
  associated with a group is estimated using a `GAP' method based on
  the luminosity of the central galaxy and its gap with other member
  galaxies. Tests using mock samples shows that this method is
  reliable, particularly for poor systems containing only a few
  members.  On average 80\% of all the groups have completeness
  $>0.8$, and about 65\% of the groups have zero contamination. Halo
  masses are estimated with a typical uncertainty $\sim 0.35\,{\rm
    dex}$. The application of the group finder to the 2MRS gives
  29,904 groups from a total of 43,246 galaxies at $z \leq 0.08$, with
  5,286 groups having two or more members.  Some basic properties of
  this group catalog is presented, and comparisons are made with other
  groups catalogs in overlap regions.  With a depth to $z\sim 0.08$
  and uniformly covering about 91\% of the whole sky, this group
  catalog provides a useful data base to study galaxies in the local
  cosmic web, and to reconstruct the mass distribution in the local
  Universe.
\end{abstract}


\section[]{Introduction}
\label{sec:intro}

One important goal in modern cosmology is to establish the
relationship between galaxies and dark matter halos in which galaxies
form and reside. Understanding this galaxy-halo connection can provide
important information about the underlying processes governing galaxy
formation and evolution.  Theoretically, there are several ways to
study this relationship. The first is to use numerical simulations
\citep{Springel2005, Wadsley2004, Bryan1995, Kravtsov2002,
  Teyssier2002, Springel2010} or semi-analytical models
\citep{vdBosch2002, Kang2005, Croton2006}. These approaches
incorporate various physical processes that are potentially important
for galaxy formation and evolution, such as gas cooling, star
formation, feedback mechanisms, and so on.  However, many processes in
such modeling have to be approximated by sub-grid implementations and
simple parameterizations, and so the results obtained are still
questionable and sometimes fail to match observations.  An alternative
method to establish the galaxy-dark matter halo connection is to adopt
an empirical approach. Models in this category includes the halo
occupation model \citep[e.g.][]{Jing1998, Peacock2000, Berlind2002,
  Zheng2005}, the conditional luminosity functions
\citep[e.g.][]{Yang2003, vdBosch2003, Yan2003, Tinker2005, Zheng2005,
  Cooray2006, vdBosch2007, Yang2012}, halo abundance matching
\citep[e.g.][]{Mo1999, Vale2004, Conroy2006, Behroozi2010, Guo2010,
  Trujillo2011}, and parametric model fitting \citep{Lu2014,
  Lu2015b}. By construction, the empirical approach can produce much
better fits to the observational data than numerical simulations and
semi-analytical models, and so the galaxy-halo relationship
established in this way is more accurate.  Yet another way of to
establish the galaxy-dark matter halo connection is to identify galaxy
systems (clusters and groups, collectively referred to as groups in
the following) to represent dark halos.  With a well-defined galaxy
group catalog, one can not only study the relationship between halos
and galaxies \citep[e.g.][]{Yang2005a, Yang2008, Lan2016,
  Erfanianfar2014, Rodriguez2015, Jiang2016}, but also investigate how
dark matter halos trace the large-scale structure of the universe
\citep[e.g.][]{Yang2005b, Yang2005c, Tal2014}.  In addition to these
statistical studies, a well-defined group sample can also be used to
reconstruct the current and initial cosmic density fields, so as to
study not only the structures but also the formation histories of the
cosmic web \citep[e.g.][]{Wang2012, Wang2013, Wang2014}.

The quality of a group sample depends on the group finder used to
identify individual groups.  During the past two decades, numerous
group catalogs have been constructed from various observations,
including the 2-degree Field Galaxy Redshift Survey (2dFGRS)
\citep{Eke2004, Yang2005a}, the DEEP2 survey \citep{Crook2007} and the
Sloan Digital Sky Survey (SDSS) \citep[e.g.][]{Berlind2006, Yang2007,
  Tago2010, Nurmi2013}.  The group finders adopted in these
investigations range from the traditional friends-of-friends (FOF)
algorithm \citep[e.g.][]{Davis1985}, to the hybrid matched filter
method \citep{Kim2002} and the ``MaxBCG" method
\citep{Koester2007}. Although the accuracy of a particular group
finder depends on the properties of the observational sample, all
group finders need to handle the same observational effects, such as
redshift distortion that impacts the clustering pattern of galaxies,
and the variations of the mean inter-galaxy separation due to apparent
magnitude limit.

In this paper, we present our construction of a galaxy group catalog
from the 2MASS Redshift Survey (2MRS), which is complete roughly to
$K_s = 11.75$ and covers 91\% of the sky \citep{Huchra2012}. Several
group catalogs have already been constructed from 2MRS.
\citet{Crook2007} constructed a group catalog using galaxies with a
magnitude limit at $K_s = 11.25$ and a FOF algorithm similar to that
of \citet{HuchraGeller1982}.  \citet{Tully2015} built a group catalog
in the volume between 3,000 and $10,000 \kms$ using a methodology
similar to that of \citet{Yang2005a}. \citet{Tempel2016} constructed a
group catalog to larger distances using a FOF algorithm.  Our goal
here is to obtain a reliable and uniform galaxy group catalog using
all galaxies in the 2MRS brighter than $K_s = 11.75$ to a redshift $z
= 0.08$. By involving a new halo mass estimation method, we are trying
to obtain a better representive halo distributions in the local
Universe.

The group finder to be used is the halo-based group finder developed
by \citet{Yang2005a}, which groups galaxies within their host dark
matter halos. This group finder is suitable to study the relation
between galaxies and dark matter haloes over a wide range of halo
masses, from rich clusters of galaxies to poor galaxy groups. It has
been tested with mock galaxy surveys, and has been applied quite
successfully to several galaxy catalogs \citep{Yang2005a,
  Weinmann2006a, Yang2007}.  The essential idea behind this group
finder is to use the relationships between halo mass and its size and
velocity dispersion when deciding the membership of a group.  Thus an
accurate estimate of the halo mass for a candidate galaxy group is a
key step.  As shown in \citet{Yang2007}, for relative deep surveys,
such as the SDSS, the group total luminosity (or stellar mass)
provides a reliable ranking of the halo mass.  In this case, halo
masses can be estimated reliably by matching the rank of the
characteristic luminosity of a group to that of halo mass given by a
halo mass function.  However, as pointed out in \citet{Lu2015}, for a
shallow survey, such as the 2MRS, where only a few bright member
galaxies in a group can be observed, the characteristic group
luminosity is no longer the best choice to estimate the halo mass
\citep[see also][for the halo mass estimation comparisons on cluster
scales]{Old2014, Old2015}.  Instead, they proposed a method that is
based on the luminosity of the central galaxy, $L_c$, and a luminosity
`GAP', $L_{gap}$, where the central galaxy is defined to be the
brightest in a group, and the luminosity gap is defined as $\log
L_{\rm gap} = \log L_c - \log L_s$, with $L_s$ being the luminosity of
the satellite galaxy of some rank (e.g. the brightest, or second
brightest satellite). The performance of the halo mass estimate is
found to be enhanced by using the `GAP' information.  Comparisons
between the true halo masses and the masses estimated with the `GAP'
method in mock catalogs show a typical dispersion of $\sim 0.3 {\rm
  dex}$.

In this paper, we modify the halo-based group finder developed by
\citet{Yang2005a} by using the `GAP' information.  The structure of
the paper is as follows.  \S\ref{sec:data} describes the samples used
in this paper, including the 2MRS galaxy sample and a mock galaxy
sample used to evaluate the performance of our group finder. In \S
\ref{sec:mass} we describe our modified halo-based group finder. The
performance of our group finder, including completeness,
contamination, purity is discussed in \S\ref{sec_test}, together with
the reconstruction of the halo mass function. In \S\ref{sec:catalog}
the properties of the group catalog constructed from 2MRS are detailed
and compared to the mock group catalog, and to the SDSS DR7 galaxy
group catalog constructed by \citet{Yang2007, Yang2012} in the
overlapping region. Finally, we summarize our results in \S
\ref{sec:summary}.  Unless stated otherwise, we adopt a $\Lambda$CDM
cosmology with parameters that are consistent with the nine-year data
release of the WMAP mission (hereafter WMAP9 cosmology): $\Omega_{\rm
  m} = 0.282$, $\Omega_{\Lambda} = 0.718$, $\Omega_{\rm b} = 0.046$,
$n_{\rm s}=0.965$, $h=H_0/(100 \kmsmpc) = 0.697$ and $\sigma_8 =
0.817$ \citep{Hinshaw2013}.

\section[]{DATA}
\label{sec:data}

\begin{figure}
\vspace{0.5cm}
\center
\includegraphics[height=7.0cm,width=7.5cm,angle=0]{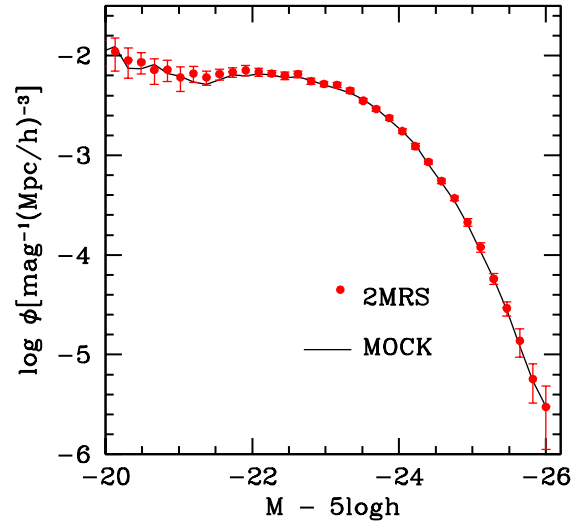}
\caption{Luminosity functions in $K_s$ band. Red points are obtained
  from the 2MRS galaxy sample, while black solid line represents the
  one obtained from the MOCK sample.  The error bars are estimated
  using 1000 boot-strap re-sampling.  }
\label{fig:lf}
\end{figure}

\begin{deluxetable*}{lcccc}
 \tabletypesize{\scriptsize} \tablecaption{Comparison between three
 samples} \tablewidth{0pt}

  \tablehead{Sample & Parent Catalog  & Group Finder   & Halo Mass & Label}  \\
  \startdata
 MOCKt  & mock 2MRS catalog &  none  &  simulation & $M_{\rm t}$\\
  MOCKg  & mock 2MRS catalog & halo-based & Gap model & $M_{\rm g}$\\
   2MRS   & 2MRS catalog & halo-based  & Gap model & $M_{\rm 2MRS}$
 \enddata
\label{tab:samples}
\end{deluxetable*}

\subsection{The 2MRS galaxy catalog}

The 2MASS Redshift Survey (2MRS) is based on the Two Micron All Sky
Survey \citep{Jarrett2000, Jarrett2003} and is complete to a limiting
magnitude of $K_s = 11.75$, and $\sim 97.6\%$ of the galaxies brighter
than the limiting magnitude have measured redshifts. The survey covers
$\sim 91\%$ of the full sky; only $\sim 9\%$ of the sky close to the
Milky Way plane is excluded \citep{Huchra2012}.  The catalog contains
about 43,533 galaxies extending out to $\sim 30,000 \kms$.  For our
analysis we only use the 43,246 galaxies with $z \leq 0.08$.  Among
these, 25 entries have negative redshifts ($-0.001 \leqslant z < 0.0$)
which are caused by the peculiar velocities of galaxies.  In our
analysis, all redshifts are corrected to the Local Group rest frame
according to \citet{Karachentsev1996}.  We also use the distance
information provided by \citet{Karachentsev2013} for some nearby
galaxies, including 22 galaxies with negative redshifts in our 2MRS
catalog, to reduce effects caused by peculiar velocities. Corrections
of Virgo infall are made to 15 galaxies in the front and back of the
Virgo cluster according to \citet{Karachentsev2014}.  Since the
redshifts of our 2MRS galaxies are low, no attempt is made to apply
any $K$- or $E$-corrections to galaxy luminosities.

From this catalog, we first measure the galaxy luminosity function
(LF) in the $K_s$ band.  We adopt the commonly used $1/V_{\rm max}$
algorithm \citep{Schmidt1968, Felten1976}, in which each galaxy is
assigned a weight given by the maximum co-moving volume within which
the galaxy could be observed.  Fig \ref{fig:lf} shows the galaxy
luminosity function so obtained from our 2MRS sample, with the error
bars estimated from 1,000 bootstrap re-samplings.  We have fitted the
LF to a Schechter function \citep{Schechter1976} and the best fit
Schechter parameters are $\log\phi^*=1.08\times 10^{-2}$,
$\alpha=-1.02$ and $M^*=-23.55$. These values are consistent with
those obtained by \citet{Crook2007} and \citet{Tully2015}.

\subsection[]{The mock 2MRS galaxy catalog}
\label{sec:mock2MRS}

\begin{figure*}
\center
\vspace{0.5cm}
\includegraphics[height=8.0cm,width=15.0cm,angle=0]{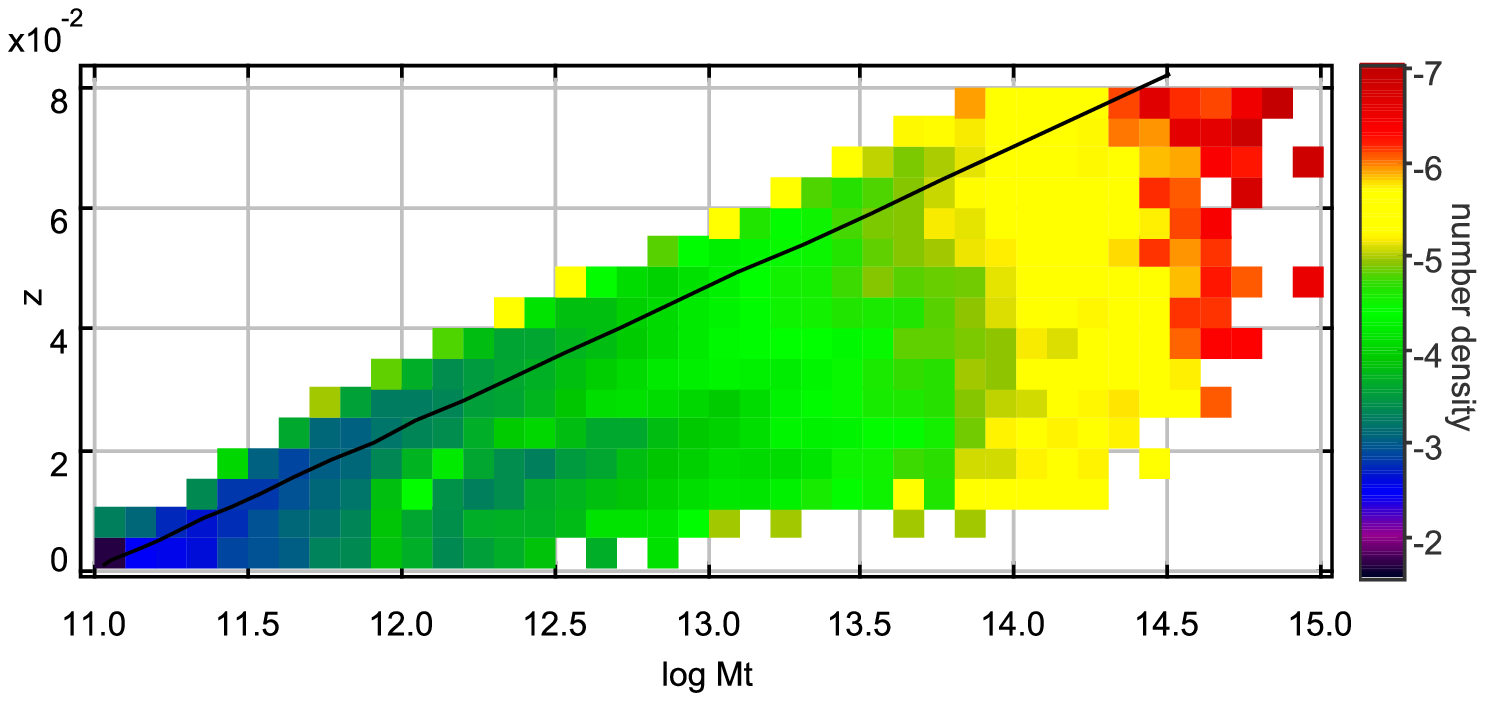}
\caption{The number densities of halos, $\log \phi [(\mpch)^{-3}]$, in
  each halo mass and redshift bins shown with color bars.  The solid
  line represents a conservative redshift limit $z_{\rm limit}=
  0.023*\log M_{\rm t} -0.26$ , below which a complete sample can be
  formed for halos with masses down to the mass given by the value of
  $M_{\rm t}$ shown by the horizontal axis. }
\label{fig:colorbar}
\end{figure*}

\begin{figure}
\center
\vspace{0.5cm}
\includegraphics[height=7.5cm,width=7.5cm,angle=0]{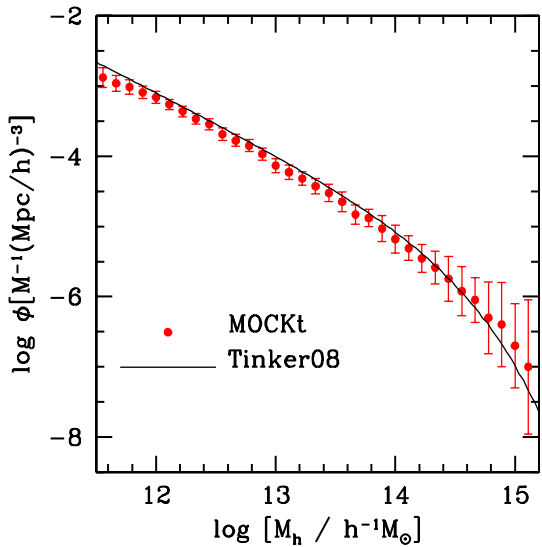}
\caption{The halo mass function of our true sample (MOCKt, dots with
  error bars).  The halo mass function given by \citet{Tinker2008} is
  also plotted in the same panel for comparison using black solid
  line.  }
\label{fig:hmf0}
\end{figure}

We construct a mock 2MRS galaxy catalog to test the performance of
our group finder and the reliability of the final galaxy group catalog.
The mock catalog is constructed as follows.

First, we use a high-resolution simulation carried out at the High
Performance Computing Center, Shanghai Jiao Tong University, using
L-GADGET, a memory-optimized version of GADGET-2
\citep{Springel2005}. A total of $3072^3$ dark matter particles were
followed in a periodic box of $500\mpch$ on a side \citep{Li2016}. The
adopted cosmological parameters are consistent with those from WMAP-9.
Each particle in the simulation has a mass of $3.4\times10^{8}
\msunh$. Dark matter halos were identified using the standard FOF
algorithm \citep{Davis1985} with a linking length of $b=0.2$ times the
mean inter particle separation.

Next, the halos are populated with galaxies of different luminosities.
We use the conditional luminosity function \citep[CLF,][]{Yang2003},
which is defined to be the average number of galaxies, as a function
of luminosity, that reside in a halo of a given mass, to link galaxies
with dark matter haloes. We make use of the set of CLF parameters
provided by \citet{Cacciato2009} to generate model galaxies with $r$
band luminosities.  Following the observational definition, the
central galaxy is defined as the brightest member and is assumed to be
located at the center of the corresponding halo. Its velocity follows
the velocity of the dark matter halo center. Other galaxies, referred
to as satellite galaxies, are distributed spherically following a NFW
\citep{NavarroFrenkWhite1997} profile where the concentration model of
\citet{Zhao2009} was adopted. Their velocities are assumed to be the
sum of the velocity of the host halo center plus a random velocity
drawn from a Gaussian distribution with dispersion given by the virial
velocity dispersion of the halo.  We refer the reader to
\citet{Lu2015} and \citet{Yang2004} for details. In general, one
  can also populate/generate mock galaxies using more sophisticated
  methods, e.g., based on sub-halos or halo merger trees, where the
  galaxies are not spherically distributed. However, as we have tested
  in \citet{Weinmann2006b}, our group finder is not very sensitive to
  the somewhat non-spherical distribution of galaxies. 

In order to convert the $r$ band magnitude to the $K_s$ band, we first
measure the {\it cumulative} luminosity function separately for both
the mock and 2MRS samples.  Assuming that galaxies more luminous in
the $r$ band are also more luminous in the $K_s$, we assign a $K_s$
band luminosity (absolute magnitude) to each galaxy. In practice, we
relate $M_r$ and $M_{K_s}$ through abundance matching:
\begin{equation}\label{eq:ab2}
\int_{-\infty}^{M_r} \phi_r (M_r') dM_r'
=
\int_{-\infty}^{M_{K_s}} \phi_{K_s} (M_{K_s}') dM_{K_s}'\,,
\end{equation}
where $\phi_r (M_r)$ and $\phi_{K_s} (M_{K_s})$ are the luminosity
functions of galaxies in $M_r$ and $M_{K_s}$, respectively.

Finally, we place a virtual observer at the center of our simulation
box and define a ($\alpha$, $\delta$)-coordinate frame, and remove all
galaxies that are located outside the survey region ($\sim 9\%$ of the
total sky). We then assign to each galaxy a redshift and an apparent
magnitude according to its distance and luminosity, and select only
galaxies that are brighter than the magnitude limit $K_s=11.75$.  Here
again, no K+E corrections are made to galaxy luminosities.  In total,
we have 41,876 galaxies in our mock 2MRS catalog. The black solid
curve in Fig. \ref{fig:lf} shows the $K_s$ band luminosity function
estimated from our mock sample.

 Apart from the luminosity function of galaxies, we set out to
  measure the true halo mass function in the mock 2MRS
  catalog. Because of the survey magnitude limit, faint galaxies
  formed in low mass halos might not be observed at high redshift,
  i.e., low mass halos can only be detected below a redshift limit.
  To properly estimate the halo mass function, one needs to have a
  complete sample of groups (halos), i.e. to obtain the limiting
  redshift for a given mass of halos, within which the selection of
  groups is complete. Unlike the luminosity for which the limiting
  redshift can be directly calculated from the magnitude limit, we use
  an empirical way to get the limiting redshift for halos. First, we
  calculate the number densities of halos in small redshift and halo
  mass bins and plot them in the $\log M_{\rm t}$-$z$ plane using
  color bars (see Fig. \ref{fig:colorbar}).  We can see that the
  number density of halos of given mass drops sharply above certain
  redshift.  Here we define the limiting redshift for a given mass
  halos as the redshift at which this rapid drop in density occurs.
  The smooth line in Fig. \ref{fig:colorbar} shows the limiting
  redshift as a function of $\log M_{\rm t}$ we use, which clearly
  represents a conservative cut to ensure completeness. Once a
  limiting redshift is adopted, we can calculate the halo mass
  function using only halos (and volume) below this
  redshift. Fig. \ref{fig:hmf0} shows the halo mass function obtained
  in this way with dots and error bars. Here again, the error bars are
  estimated using 1000 bootstrap re-samplings. For comparison, we also
  show, using the solid line, the theoretical model predictions given
  by \citet{Tinker2008}.  Compared with the model prediction, the data
  points are slightly lower at intermediate to low mass range, which
  is mainly due to cosmic variance, since the overall halo mass
  function in the whole simulation box is quite consistent with
  theoretical predictions \citep[see][]{Li2016}.

Since we have both the true halo and the galaxy membership
informations in our mock 2MRS sample, we can use it to test the
performance of our group finder. Together with the 2MRS
  observational sample, there are three galaxy group catalogs involved
  in this paper. We refer to the two catalogs related with the mock
  2MRS samples as `MOCKt' and `MOCKg' respectively. The former catalog
  indicates the true group memberships in the FOF dark matter halos
  obtained directly from the simulation, where we use $M_{\rm t}$ to
  represent the {\it true} halo mass.  The latter is constructed using
  our group finder, where the halo masses are estimated using our
  `GAP' related mass estimator and are named as $M_{\rm g}$.  Finally,
  the group catalog constructed from the 2MRS data is referred to as
  `2MRS' and the related halo mass are named as $M_{\rm 2MRS}$ .  For
  clarity, we list the differences of the three definitions in
  Table. \ref{tab:samples}, including the group finders that were used
  to identify galaxy groups, and the methods used to estimate the halo
  masses.  Apart from the above three specific halo mass definitions,
  we use $M_h$ to represent the general halo masses, including those
  used in theoretical model predictions \citep[e.g.][]{Tinker2008}.

\section{The modified halo-based group finder}
\label{sec:mass}

One of the key steps in the halo-based group finder
\citep{Yang2005a,Yang2007} is to have accurate estimates of the halo
masses of candidate galaxy groups. As demonstrated in
\citet{Yang2007}, halo mass is tightly correlated with the total
luminosity of member galaxies. In practice, however, one can only
estimate a characteristic luminosity which is the sum of the
luminosities of member galaxies brighter than some given limit.  For a
relatively deep survey such as the SDSS, where the limit can be set
sufficiently low, the characteristic luminosity is a good proxy of the
total luminosity and so can be used to indicate halo mass.  For a
shallow survey like the 2MRS, on the other hand, only a few (in most
cases one or two) brightest member galaxies in the halos can be
observed. The characteristic luminosity is no longer the best halo
mass estimator, and an alternative is needed. In this paper, we
implement the `GAP' method proposed by \citet{Lu2015}.

\subsection[]{The GAP halo mass estimator}
\label{sec:gap}

\begin{figure}
\center
\vspace{0.5cm}
\includegraphics[height=8.0cm,width=9.0cm,angle=0]{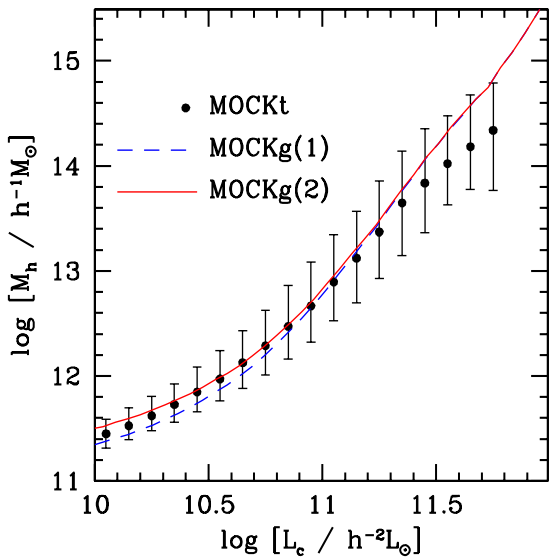}
\caption{$L_c-M_h(M_{\rm g})$ relation given by the MOCKg sample
  using abundance matching between the cumulative luminosity function
  of galaxies and the halo mass function. `Round 1' relation is
  obtained using all mock galaxies [blue dashed line, labelled as
  MOCKg(1)] while `Round 2' is obtained using central galaxies only
  [red line labelled as MOCKg(2)].  The true $L_c-M_h(M_{\rm t})$
  relation given by the MOCKt sample is plotted with black solid
  points with error bars which indicate the $16\%-84\%$ percentiles of
  the distributions around the median values.}
\label{fig:MhLcmock}
\end{figure}

\begin{figure*}
\center
\vspace{0.5cm}
\includegraphics[height=10.5cm,width=11.5cm,angle=0]{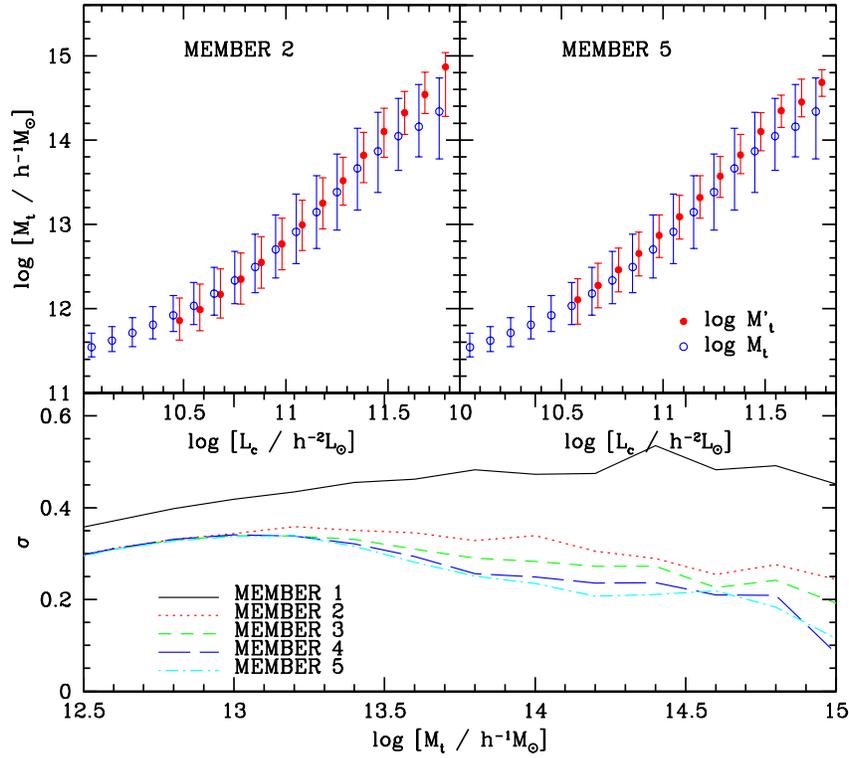}
\caption{Comparison between the original (open blue circles) $\log
  M_{\rm t}$ and corrected (solid circles) $\log M'_{\rm t}$ halo
  masses by using the luminosity gap between the central (brightest)
  and the second brightest (top left) and the fifth brightest member
  galaxies (top right), respectively. The error bars indicate the
    $16\%-84\%$ percentiles of the distributions. The standard
  variances $\sigma$ between estimated halo mass and the true halo
  mass are illustrated in the bottom panel. As the legend indicates,
  results are shown for groups with 2, 3, 4 and 5 members, while the
  halo masses estimated only by using central galaxies (member 1) are
  also given in the same panel.  }
\label{fig:M_correct}
\end{figure*}

\begin{deluxetable*}{lccccc}
  \tabletypesize{\scriptsize} \tablecaption{Parameters of the $\Delta
    \log M_g$ model obtained from mock 2MRS sample. [See
    Eqs. (\ref{eq:DM_func}) \& (\ref{eq:eta_abc})]} \tablewidth{0pt}

  \tablehead{$\Delta \log M_g$ & $\beta_1$  & $\alpha_2$   & $\beta_2$   &
    $\beta_3$   & $\gamma_3$ } \\
  \startdata

  MEMBER 2 & $    10.81^{+ 0.18}_{- 0.19}$ &  $     0.36^{+ 1.60}_{- 0.26}$ &  $   -15.44^{
+ 3.35}_{- 7.86}$ &  $    10.39^{+ 0.11}_{- 0.24}$ &  $     1.94^{+ 0.83}_{- 0.4
1}$    \\
   \\
  MEMBER 3 & $    10.21^{+ 0.39}_{- 0.10}$ &  $     0.23^{+ 0.54}_{- 0.14}$ &  $   -13.40^{
+ 1.21}_{- 3.97}$ &  $     9.90^{+ 0.36}_{- 0.10}$ &  $     2.21^{+ 0.32}_{- 0.2
7}$    \\
   \\
  MEMBER 4 & $    9.98^{+ 0.32}_{- 0.18}$ &  $     0.20^{+ 0.25}_{- 0.09}$ &  $   -13.39^{
+ 1.10}_{- 2.76}$ &  $     9.81^{+ 0.30}_{- 0.17}$ &  $     2.45^{+ 0.24}_{- 0.1
5}$     \\
    \\
  MEMBER 5 & $    9.77^{+ 0.33}_{- 0.07}$ &  $     0.13^{+ 0.15}_{- 0.01}$ &  $   -13.67^{
+ 1.18}_{- 0.94}$ &  $     9.67^{+ 0.28}_{- 0.07}$ &  $     2.54^{+ 0.15}_{- 0.0
8}$      \\

 \enddata
\label{tab:DeltaMmock}
\end{deluxetable*}

In the `GAP' method, one first needs to estimate the $L_c$-$M_h$
relation. For MOCKt samples, since every groups have the true
  central galaxies and true halo masses from the simulation, we can
  obtain this relation directly.  Hereafter we refer the $L_c-M_h$
  relation obtained directly from the simulation as the intrinsic
  (true) relation. On the other hand, for observational samples, one
can obtain this relation from the conditional luminosity function
model \citep[e.g. ][]{Yang2003} or from halo abundance matching
\citep[e.g.][]{Mo1999,ValeOstriker2006,Conroy2006, Behroozi2010,
  Guo2010}.  Here we adopt the latter and assume that there is a
monotonic relation between the luminosity of central galaxy and the
mass of dark matter halo, so that a more luminous galaxy resides a
more massive halo. We can then get an initial estimate of the dark
matter halo mass for each central galaxy from
\begin{equation}\label{eq:ab3}
  \int_{L_c}^\infty n_c (L_c') dL_c' =
  \int_{M_h}^\infty n_h (M_h') dM_h'\,,
\end{equation}
where, $n_c (L_c)$ is the number density of central galaxies with
luminosity $L_c$ and $n_h (M_h)$ is the number density of halos
  (or halo mass function) with mass $M_h$. In this paper, we adopt
  theoretical halo mass function given by \citet{Tinker2008}. Note
that, in this abundance matching approach, we also need to know
whether a galaxy is a central or a satellite. Since we are trying to
find galaxy groups within the observation (the 2MRS in our case), we
can easily separate galaxies into centrals and satellites with the
help of group memberships. As we will show later, although the
$L_c-M_h$ relation we obtain may deviates from the true one,
especially at the massive end, the deviation can be compensated to
some extent by our `GAP'-based correction factor.

Our modeling of the $L_c-M_h$ relation using Eq. (\ref{eq:ab3}) is
carried out via the following two steps. First, before we are able to
separate galaxies into centrals and satellites with the help of group
memberships, we assume that all of them are centrals \citep[as shown
in ][ more than 60\% of the galaxies are centrals]{Yang2008}.  To
  show the performance, We have applied this to our mock 2MRS sample,
and obtain the `Round 1' $L_c-M_{\rm g}$ relation, which is shown
  in Fig. \ref{fig:MhLcmock} as the blue dashed line.

For comparison, we also plot, as black solid points, the true median
$L_c-M_{\rm t}$ relation obtained from the true centrals and true
  halo masses in the simulation, with error bars indicate the
  $16\%-84\%$ percentiles of the distributions.  Compared to the true
relation, we see that, the Round 1 relationship shows a general
agreement with the true one, with a slight over-prediction of
the halo masses at the bright end and slight under-prediction at the
faint end. The deviation at the massive end is caused by the Malmquist
bias in the $L_c-M_{\rm g}$ relation which can be corrected by the
`GAP' \citep[see ][]{Lu2015}.  The deviation at the faint end is
caused by the inclusion of all the galaxies (including satellites) in
our abundance matching. As we apply our group finder to the galaxy
catalog in the next step, the group membership will enable us to
separate galaxies into centrals and satellites. We can then limit the
application of the abundance matching to centrals only, and improve
the $L_c-M_h$ relation. After two to three iterations we converge to a
new set of group memberships and a new $L_c-M_{\rm g}$ relationship,
which is referred to as 'Round 2' and shown as the solid red line in
Fig. \ref{fig:MhLcmock}.  After this step, there is no longer any
systematic deviation of the $L_c-M_{\rm g}$ relationship relative to
the true one at the low mass end.

With the $L_c-M_{\rm g}$ obtained in this step, we can estimate the
`luminosity gap', which is defined as the luminosity ratio between the
central and a satellite galaxy in the same halo, $\log L_{\rm gap}=
\log(L_c/L_s)$ \citep[see][]{Lu2015}.  The halo mass is then estimated
using the relation,
\begin{equation}\label{eq:Mfunc}
\log M_{\rm g}(L_c,L_{\rm gap}) = \log M_{\rm g}(L_c) + \Delta \log
M_{\rm g}(L_c,L_{\rm gap}) \,.
\end{equation}
This halo mass estimator consists of two parts. The first part
  is an empirical relation between $M_{\rm g}$ and $L_c$ derived from
  Eq. (\ref{eq:ab3}) which is represented by the first term on the
  right side. Another part is the amount of correction to that
  relation, which is represented by the second term $\Delta \log
  M_{\rm g}(L_c,L_{\rm gap})$. In order to model this correction term,
  we use the following functional form,
\begin{equation}\label{eq:DM_func}
  \Delta \log M_{\rm g}(L_c,L_{\rm gap}) = \eta_a \exp(\eta_b \log L_{\rm gap}) + \eta_c\,.
\label{eq:delM}
\end{equation}
The parameters $\eta_a$, $\eta_b$ and $\eta_c$ all depend on $L_c$ as:
\begin{eqnarray}\label{eq:eta_abc}
\eta_a(L_c)~ &=&~ \exp(\log L_c-\beta_1) \nonumber \\
\eta_b(L_c)~ &=&~ \alpha_2(\log L_c +\beta_2)  \\
\eta_c(L_c)~ &=&~ -(\log L_c - \beta_3)^{\gamma_3} \nonumber
\end{eqnarray}
which is specified by five free parameters.

For a given $L_c - M_{\rm g}$ relation, we fit the model to the true
halo masses $M_{\rm t}$ of our galaxy systems (groups) in our mock
sample to have the minimum variances between $\log M_{\rm t}$ and
$\log M_{\rm g}(L_c,L_{\rm gap})$ \citep[see][for details]{Lu2015}.
Table \ref{tab:DeltaMmock} presents the set of best fit values of
these parameters.  Since the (mock) 2MRS sample is shallow, we provide
the parameters up to 5 group members.  As an illustration,
Fig. \ref{fig:M_correct} shows the performance of this halo mass
estimator. In the top two panels, the original $L_c-M_t$ relations are
shown as the open circles; the GAP-corrected relations are shown as
the solid points, with the left panel showing results for $L_s=L_2$
and the right for $L_s=L_5$. To see the improvement, we define a
  `pre-corrected' halo mass,
\begin{equation}\label{eq:M'_h}
\log M'_{\rm t} = \log M_{\rm t} - \Delta \log M_g(L_c,L_{\rm gap})\,.
\end{equation}
If the correction term $\Delta \log M_{\rm g}(L_c,L_{\rm gap})$ can
perfectly describe the scatter in the original relation $L_c-M_{\rm
  t}$, then there would be no scatter in the $L_c-M'_{\rm t}$. We can
see that, the scatter in the $L_c-M'_{\rm t}$ is significantly reduced
compare to that in the $L_c-M_{\rm t}$ relation. For massive
halos/groups, this improvement is quite notable where the scatter is
reduced almost by a factor of two.  The bottom panel of
Fig. \ref{fig:M_correct} shows the standard deviation $\sigma$ of the
halo mass $\log M_{\rm g}(L_c,L_{\rm gap})$ obtained by
Eq.(\ref{eq:Mfunc}) from the true halo mass $\log M_{\rm t}$. In both
\citet{Lu2015} and this paper, we find that using $L_5$ gives the best
correction to the halo mass.  As shown \citet{Lu2015}, such a
correction factor is quite independent of the galaxy formation model
used to construct the mock catalog.  In this paper we use the set of
best fit parameters only up to the fifth ranked member (see below).

\subsection[]{The Group Finder}
\label{sec:gf}

The group finder adopted here is similar to that developed by
\citet{Yang2005a}.  It uses the general properties of dark matter
haloes, namely size and velocity dispersion, to iteratively find
galaxy groups.  Tests show that this group finder is powerful in
linking galaxies with dark matter halos, even in the case of single
member groups.  As we pointed out earlier, the halo mass estimation
adopted in \citet{Yang2005a,Yang2007} is based on the ranking of the
characteristic group luminosity and proves to be quite reliable for
surveys like the 2dFGRS and SDSS. For the 2MRS considered here, we use
the `GAP'-corrected estimator described above.  The modified group
finder with this halo mass estimator consists of the following main
steps:

~~\\
\textbf{Step 1: Start the halo-based group finder.}
\vspace{3mm}

In the earlier version of the halo-based group finder, the first step
is to use the FOF algorithm \citep{Davis1985} with very small linking
lengths in redshift space to find potential groups.  Here we assume
all galaxies in our catalog are candidate groups.  The halo mass of
each candidate group is calculated using the $L_c-M_h$ relation
obtained in Eq. (\ref{eq:ab3}) (Round 1).

~~\\
\textbf{Step 2: Update group memberships using halo information.}
\vspace{3mm}

After assigning halo masses to all the candidates, groups are sorted
according to their halo masses. Starting from the most massive one, we
estimate the size and velocity dispersion of the dark matter halo,
using the halo mass currently assigned to it.  A dark matter halo is
defined to have an over-density of 180.  For the WMAP9 cosmology
adopted here, the radius is approximately
\begin{equation}
r_{180} = 1.33 \mpch \left( \frac{M_h}{10^{14}\msunh}\right)^{1/3}
\left(1+z_{\rm group}\right)^{-1},
\end{equation}
Here, $z_{\rm group}$ is the redshift of the group center.  The
line-of-sight velocity dispersion of the halo is
\begin{equation}\label{eq:sigma}
  \sigma = 418\kms\left( \frac{M_h}{10^{14}\msunh}\right)^{0.3367}\,.
\end{equation}
Finally, following \citet[hereafter Y07]{Yang2007}, we use the
luminosity weighted center of member galaxies as the new group center.

 Then, one can assign new member galaxies to the group according
  to the tentative group center, tentative estimates of halo size and
  velocity dispersion obtained in the above steps. The phase-space
  distribution of galaxies is assumed to follow that of dark matter,
  and the group center is assumed to coincide with the center of
  halo. We use the following function of the projected distance $R$
  and $\Delta z = z - z_{\rm group}$ to describe the number density of
  galaxies at $z$ in the redshift space around the group center at
  redshift $z_{\rm group}$:
\begin{equation}
P_M(R,\Delta z) = {H_0\over c} {\Sigma(R)\over {\bar \rho}} p(\Delta z) \,,
\end{equation}
where $c$ is the speed of light and $\bar{\rho}$ is the average
density of the Universe. We assume the projected surface density,
$\Sigma(R)$, is given by a (spherical) NFW
\citep{NavarroFrenkWhite1997} profile:
\begin{equation}
\Sigma(R)= 2~r_s~\bar{\delta}~\bar{\rho}~{f(R/r_s)}\,,
\end{equation}
where $r_s$ is the scale radius, and the shape function is
\begin{equation}
\label{fx}
f(x) = \left\{
\begin{array}{lll}
\frac{1}{x^{2}-1}\left(1-\frac{{\ln
{\frac{1+\sqrt{1-x^2}}{x}}}}{\sqrt{1-x^{2}}}\right)   &  \mbox{if   $x<1$}  \\
\frac{1}{3}   &   \mbox{if   $x=1$}   \\
\frac{1}{x^{2}-1}\left(1-\frac{{\rm
      atan}\sqrt{x^2-1}}{\sqrt{x^{2}-1}}\right) & \mbox{if $x>1$}
\end{array} \right.\,.
\end{equation}
The normalization of the profile depends on the concentration
$c_{180}=r_{180}/r_s$ as:
\begin{equation}
\bar{\delta} = {180 \over 3} {c_{180}^3 \over {\rm ln}(1 + c_{180}) -
c_{180}/(1+c_{180})} \,,
\end{equation}
where the concentration model of \citet{Zhao2009} is adopted. 
The redshift distribution of galaxies within the halo is assumed
to have a normal distribution, and can be described as follows,
\begin{equation}
p(\Delta z)=  {1 \over  \sqrt{2\pi}} {c \over  \sigma (1+z_{\rm  group})} \exp
\left [ \frac {-(c\Delta z)^2} {2\sigma^2(1+z_{\rm group})^2}\right ] \,,
\end{equation}
where $\sigma$ is the rest-frame velocity dispersion given by
equation~(\ref{eq:sigma}).  So defined, the three-dimensional
  density in redshift space is $P_M(R,\Delta z)$. Then, we apply the
  following procedures to assign a galaxy to a particular group. For
each galaxy we loop over all groups, and compute the distances $R$ and
$\Delta z$ between the galaxy and the group center. An appropriately
chosen background level $B=10$ is applied to the density contrast for
galaxies to be assigned to a group. If, according to this criterion, a
galaxy can be assigned to more than one group it is only assigned to
the one with the highest $P_M(R,\Delta z)$.  Finally, if all members
of two groups can be assigned to one, they are merged into a single
group. Note that in our group finder, the background
  parameter $B=10$ is set to ensure the balance between the
  interlopers and completeness of group memberships. A lower $B$ value
  will increase both the completeness of the group memberships and the
  number of interlopers, especially in massive groups.  Thus for those
  who care most about the completeness of the group memberships only,
  a lower value of $B$ (e.g. $B=5$) can be used \citep[see
  ][]{Yang2005a}. 

~~\\
\textbf{Step 3: Update halo mass with `GAP' correction.}
\vspace{3mm}

Once the new membership to a group is obtained, we use the new central
and satellite galaxy system to estimate the halo mass using the `GAP'
method described by Eq. (\ref{eq:Mfunc}).  For each candidate group,
we use the $L_c-M_h$ relation and the luminosity gap $\log L_{\rm
  gap}$ between the central galaxy and the faintest satellite (if the
group contains less than 5 members) or the fifth brightest galaxy (if
the group has membership equal to or larger than 5), to estimate the
halo mass.  In practice, we only apply the luminosity gap correction
for centrals in the luminosity range $10.5 \leq \log L_c \leq
11.7$. As shown in the top panels of Fig. \ref{fig:M_correct}, fainter
($\log L_c \leq 10.5$) central galaxies are basically isolated. For
$\log L_c \geq 11.7$, we found that using the value of $L_c$ directly
in the GAP leads to over-correlation. Thus, for these systems we set
$\log L_c = 11.7$ to estimate the GAP correction. In addition, since
our galaxy sample is magnitude limited to $K_s = 11.75$, our method
also suffers from a `missing satellite' problem, in that some groups
do not contain any satellites brighter than the magnitude limit. As an
attempt to partly correct for this, we assume that each galaxy group
that contains only one member (a central) has a potential member
satellite galaxy with an apparent magnitude $K_s = 11.75$, which
corresponds to a limiting luminosity $L_{\rm limit}$ at the distance
of the group.  A `GAP' correction, $\log L_c - \log L_{\rm limit}$, is
also applied to all groups of single membership with $\log L_c - \log
L_{\rm limit}\geq 0.5$, and the final halo mass of such a group is set
to be the average value between this mass and the original mass based
on the central galaxy alone.

~~\\
\textbf{Step 4: Update the $L_c - M_h$ relation and Iterate.}
\vspace{3mm}

Once all the groups have been updated for new memberships, we can
distinguish between centrals and satellites. We use the updated
central galaxy sample to update the $L_c-M_h$ relation (Round 2) to be
used to assign halo masses to tentative groups.  We iterate Steps 2-4
until convergence is reached. Typically three iterations are needed to
achieve convergence. Our final catalog is the collection of all the
converged groups with information about their positions, galaxy
memberships, and halo masses.

~~\\
\textbf{Step 5: Update the final halo masses of groups.}
\vspace{3mm}

Once all the groups (memberships) have been finalized, we perform a
final update of the halo masses of groups using an abundance matching
method so that the halo mass function of the groups is consistent with
theoretical predictions \citep[e.g.][]{Yang2007}.  In performing the
halo abundance matching, we measure the {\it cumulative} halo mass
functions of groups following the procedures described in section
\ref{sec:mock2MRS}.

\section{Test with Mock catalogs}
\label{sec_test}

In this section, we test the performances of our group finder, both in
halo masses and group memberships it assigns, by comparing the groups
selected by our group finder (MOCKg) with the true groups defined
  by simulation (MOCKt) in our mock 2MRS sample. 

\subsection{Completeness, Contamination and Purity}
\label{sec:compl}

\begin{figure}
\center
\vspace{0.5cm}
\includegraphics[height=16.0cm,width=8.0cm,angle=0]{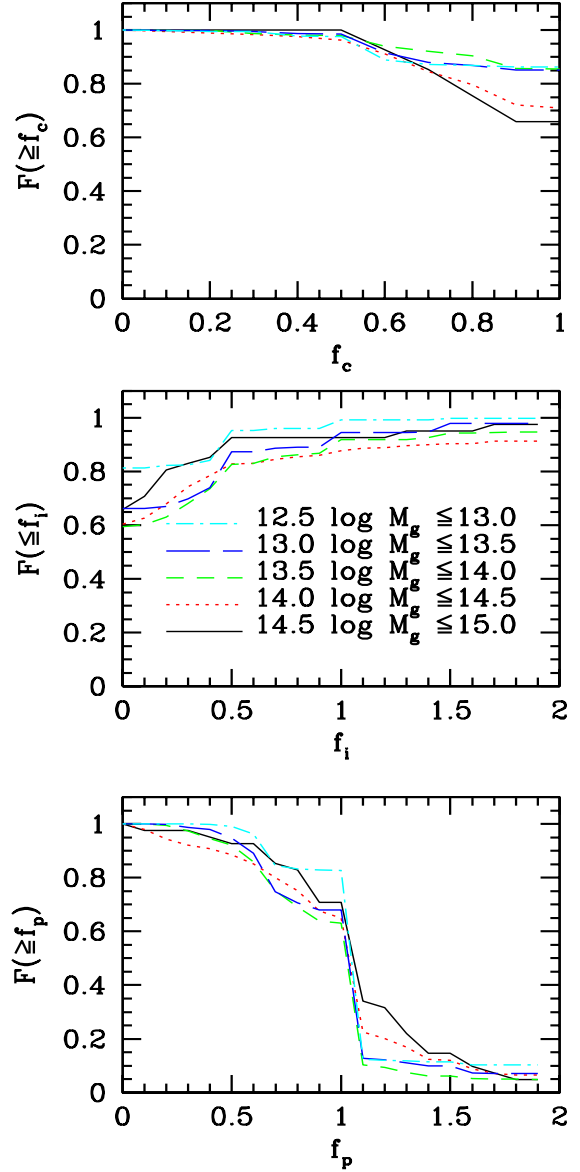}
\caption{The top, middle and bottom panels show the cumulative
  distributions of completeness, $f_c$ (the fraction of true members),
  contamination, $f_i$ (the fraction of interlopers), and purity,
  $f_p$ (ratio between the number of true members and the total number
  of group members).  These values are number weighted. Different lines
  represent the results for groups in halos of different masses, as
  indicated. Results are plotted for groups with at least 2 members,
  since groups with only 1 member have, by definition, $f_i = 0$.}
\label{fig:compl}
\end{figure}

\begin{figure*}
\vspace{0.5cm}
\center
\includegraphics[height=5.5cm,width=15.5cm,angle=0]{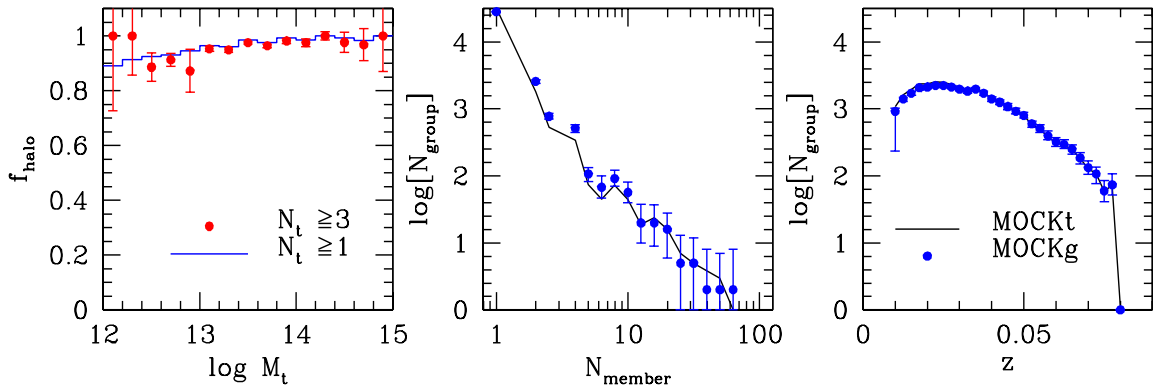}
\caption{Left panel: global completeness $f_{halo}$, defined as the
  fraction of halos in the mock sample whose brightest member has
  actually been identified as the brightest (central) galaxy of its
  group, as function of the true halo mass $M_{\rm t}$.  Results are
  shown for all halos (solid blue line) and for those with at least
  three members in the mock sample (dashed red line).  Middle and
  right panels: the distributions of richness (middle) and redshift
  (right) of groups. The results obtained from the group catalog
  constructed from the mock galaxy sample (MOCKg) are represented by
  solid points. The distributions obtained from true sample (MOCKt)
  are shown by the solid lines. All the error bars shown in the figure
  are estimated from 1000 bootstrap re-samplings. }
\label{fig:fhalo}
\end{figure*}

Starting from a total of 41,876 galaxies in the mock 2MRS
  sample, our group finder returns 32,368 galaxy groups, among which
  4,225 have 2 or more members, and the rest only one member. 
 We follow \citet{Yang2005a,Yang2007} to assess the performance of
  the group finder. The procedure is as follows. First, for each
  group, $k$, in the MOCKg sample, we identify the halo with ID,
  $h_k$, in MOCKt according to its brightest member. Then, we define
  the total number of true members belong to halo $h_k$ to be $N_t$.
  Among $N_t$, the number of true members that belong to the group $k$
  is written as $N_s$. The number of interlopers (group members that
  belong to a different halo) in the group $k$ is defined as $N_i$,
  while the total number of group members selected by our group finder
  in MOCKg is assumed to be $N_g$, and $N_g = N_i + N_s$. If our group
  finder is perfect, it should have $N_i = 0$ and $N_t = N_s =
  N_g$. With these numbers, we can define the following three
  quantities:
\begin{itemize}
\item COMPLETENESS: $f_c \equiv N_s/N_t$;
\item CONTAMINATION: $f_i \equiv N_i/N_t$;
\item PURITY: $f_p \equiv N_t/N_g$\,.
\end{itemize}
Here, $f_p = 1/(f_c+f_i)$. If the group is incomplete, then the
  COMPLETENESS $f_c < 1$.  For the CONTAMINATION $f_i$, it can be
  larger than unity.  Finally, for PURITY $f_p$, when the number of
  interlopers is larger than the number of missed true members, $f_p <
  1$, on the other hand, if the number of missed true members is lager
  than the interlopers, then $f_p > 1$. If our group finder is
  perfect, then it should have $f_c =f_p =1$ and $f_i=0$ for all the
  groups.  Note also that the value for the background level $B=10$
was tuned to maximize the average value of $f_{\rm c} (1 - f_{\rm
  i})$, as described in \citet{Yang2005a}.

Fig. \ref{fig:compl} shows the reliability of our group
  catalog constructed from the mock 2MRS galaxy sample. Following
Y07, here, only groups in MOCKg with richness $N_g \geq 2$ are
  included since the single groups with only one member always have
  zero contamination $f_i=0$ as defined above. The upper panel shows
  the cumulative distributions of the COMPLETENESS $f_c$.  The groups
  with different true halo mass are represented with different lines
  as indicated. The fractions of groups with 100 percent completeness
  ($f_c = 1$) range from $\sim 85\%$ to $\sim65\%$ depends on halo
  mass, which shows that more massive groups tend to have lower
  completeness fraction. Since the massive groups with larger velocity
  dispersions tend to have larger $f_i$ due to contamination of
  foreground and background galaxies, meanwhile, the purpose of our
  group finder is to maximizing the average value of $f_c(1-f_i)$, a
  compromise between $f_c$ and $f_i$, a background level $B=10$ is
  thus chosen.  A smaller $B$ value will increase both $f_c$ and $f_i$
  values in more massive groups, which is not preferably adopted in
  our investigation. Overall, more than 90\% of our groups have
  COMPLETENESS $f_c > 0.6$. For groups with $\log M_{\rm g} \le
14.0$, about 80\% of all groups have $f_c >0.8$; only for massive
halos with $\log M_{\rm g} > 14.0$ is this fraction a little lower,
$\sim 75-80\%$.

The middle panel of Fig. \ref{fig:compl} shows the cumulative
distribution of the CONTAMINATION $f_i$. The fraction of groups with
$f_i = 0$ ranges from 60\% to 80\%, depending on the halo mass, while
$\sim 85\%$ of all the groups have $f_i< 0.5$. The interlopers
producing the contamination are either nearby field galaxies or the
member galaxies of nearby massive groups, especially for systems that
are along the same line of sight.  Although the results for different
halo masses are similar, groups in the lowest mass bin seems to have
the highest fraction of interlopers.

Finally, the cumulative distribution of the PURITY $f_p$ is shown
  in the lower panel. On average, the number of groups which have $f_p
  < 1$ is about the same with that have $f_p > 1$. The break at
  $f_p=1$ indicates that the number of recovered group members is
  about the same as the number of the true members. Thus, the sharper
  the break is, the better. The ideal case, if our group finder is
  perfect, it should be a step function at $f_p = 1$. As one can
see, only for massive haloes there is a small fraction, $\sim 10\%$,
with $f_p < 0.5$, and a significant fraction, $\sim15\%$, with $f_p >
1.5$.

We also calculate the COMPLETENESS, CONTAMINATION and PURITY in
  terms of the total luminosity rather than the number of member
  galaxies as shown in Fig. \ref{fig:compl}.  Although not explicitly
  shown here, the corresponding results are very similar to those in
  terms of the number of member galaxies.

We now turn to the global properties of groups. First, we
  examine the \textsl{global completeness}, $f_{halo}$ which defined
  to be the fraction of halos in the MOCKt whose brightest members
  have actually been identified as the brightest (central) galaxies of
  the corresponding groups in the MOCKg. The left panel of
Fig. \ref{fig:fhalo} shows $f_{halo}$ as a function of the true halo
mass, obtained from our MOCKt sample for halos with $N_t \geq 1$
(solid blue line) and $N_t \geq 3$ (dashed red line), respectively.
As one can see, more than 90\% of all the true halos with masses
  $\geq 10^{13}h^{-1}M_{\bigodot}$ are selected by our group finder,
  almost independent of their richness.  There is a weak trend with
halo mass, in the sense that the performance of the group finder, in
terms of $f_{halo}$, is better for more massive halos.  The other
global properties we examined are the richness and redshift
distributions of galaxy groups. Shown in the middle and right panel of
Fig. \ref{fig:fhalo} are the two resulting distributions for the MOCKg
and MOCKt catalogs, respectively, and good agreement is clearly seen
between MOCKg and MOCKt.

\subsection{Halo masses of galaxy groups}
\label{sec:mockhmf}

\begin{figure*}
\center
\vspace{0.5cm}
\includegraphics[height=8.5cm,width=13.0cm,angle=0]{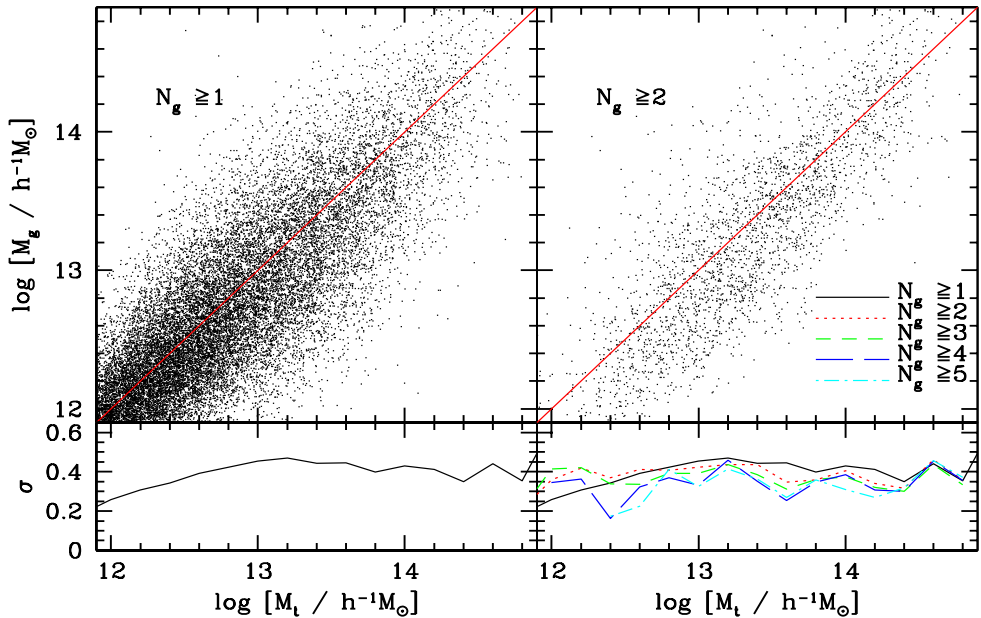}
\caption{Comparison between the estimated halo mass $M_{\rm g}$ by
  group finder and the true halo mass $M_{\rm t}$ in the simulation.
  Points in top panels are shown for all groups (left) and groups with
  more than one members (right). Red lines show the relation $M_{\rm
    g}=M_{\rm t}$. The standard variations from the red lines are
  shown as $\sigma$ in the bottom small panels, with different lines
  represent results for groups with different richness, as indicated.
}
\label{fig:Mhtestmock}
\end{figure*}

An important aspect of our group finder is the assignment of halo
masses to the groups. An accurate halo mass estimate is not only
important in determining group memberships according to common dark
halos, but also in the applications of our group catalog to the
investigations of galaxy populations in halos and large-scale
structure traced by galaxy groups.  As described above, our halo mass
estimate is based on the ranking of the `GAP' corrected luminosities
of central galaxies, and our test in \S \ref{sec:gap} using true halo
masses and group membership information in MOCKt shows that this halo
mass estimate is unbiased and has scatter typically of 0.35 to 0.2 dex
for halos with masses between $\sim 10^{13}$ to $\sim
10^{15}\msunh$. However, in real applications, the halo mass estimate
also suffers from survey selection effects, contamination and
incompleteness of group memberships, and so on. The accuracy of the
mass estimate is expected to be reduced.  Here we check the accuracy
of our halo mass estimates in MOCKg catalog which is
  constructed from the mock 2MRS sample using our group finder.

Fig. \ref{fig:Mhtestmock} shows the comparison between the true halo
mass $M_{\rm t}$ and the estimated halo mass $M_{\rm g}$ from the
galaxy group catalog we constructed from the mock 2MRS
  sample.  An estimated group in MOCKg is paired with a true one in
MOCKt if they both contains the same central galaxy, and we compare
the halo mass assigned by our group finder in MOCKg with the true halo
mass in MOCKt. Note that because of the contamination of our
  group finder (merger or fragmentation), only about $\ga 90\%$ groups
  are paired and shown here (cf. the left panel of
  Fig. \ref{fig:fhalo}). The left panel of Fig. \ref{fig:Mhtestmock}
shows the comparison for all groups while the right panels for MOCKg
groups with more than one member $N_g\ge 1$.  The corresponding
standard deviations are plotted in the bottom two panels, with
different lines representing groups of different richness.
Fig. \ref{fig:Mhtestmock} shows that the deviation is typically
between 0.2 - 0.45 dex for all groups, with some dependence on halo
mass. For $N_g\ge 1$, the scatter appears to be the largest for halos
with $M_{\rm t}\sim 10^{13}\msunh$.  The bottom right panel shows that
the mass estimate is improved as the group richness increases.  For
groups with $N \geq 3$, the scatter is about 0.35 dex, which is
comparable to that obtained by Y07 for SDSS groups.

\begin{figure}
\vspace{0.5cm}
\center
\includegraphics[height=7.5cm,width=7.5cm,angle=0]{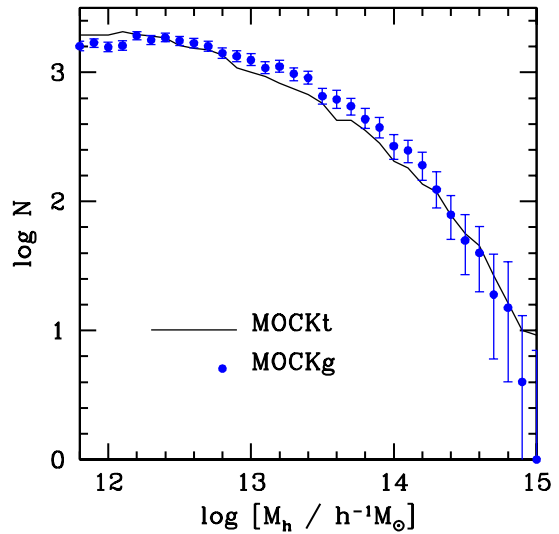}
\caption{The number distribution of groups/halos as a function of halo
  mass. The results obtained from the group catalog constructed from
  the mock galaxy sample (MOCKg) are represented by solid points. The
  error bars are estimated from 1000 bootstrap re-samplings. The black
  solid curve represents the results obtained using the true halos in
  the mock 2MRS sample (MOCKt).  }
\label{fig:hmf}
\end{figure}

The number distribution of groups as a function of halo mass
  recovered is another important test of the group finder.  In
  Fig. \ref{fig:hmf} we show, as the solid points with error bars
  (obtained by 1000 bootstrap re-samplings), the number distribution
  of groups as a function of halo mass obtained from the MOCKg catalog
  selected by our group finder.  For comparison, the distribution of
  true halos given by the MOCKt catalog is shown as the solid
  line. We see that the number distribution obtained with our group
  finder matches fairly well with the true halos. The slight over
  prediction of the number of groups at the intermediate mass range in
  the MOCKg is caused by the fact that we have forced the final halo
  mass function of groups to agree with theoretical model prediction,
  while the real mass function of MOCKt may deviate from the
  theoretical prediction due to cosmic variance
  (cf. Fig. \ref{fig:hmf0}).

\section{The 2MRS Galaxy Group Catalog}
\label{sec:catalog}

\begin{figure*}
\center
\vspace{0.5cm}
\includegraphics[height=7.0cm,width=14cm,angle=0]{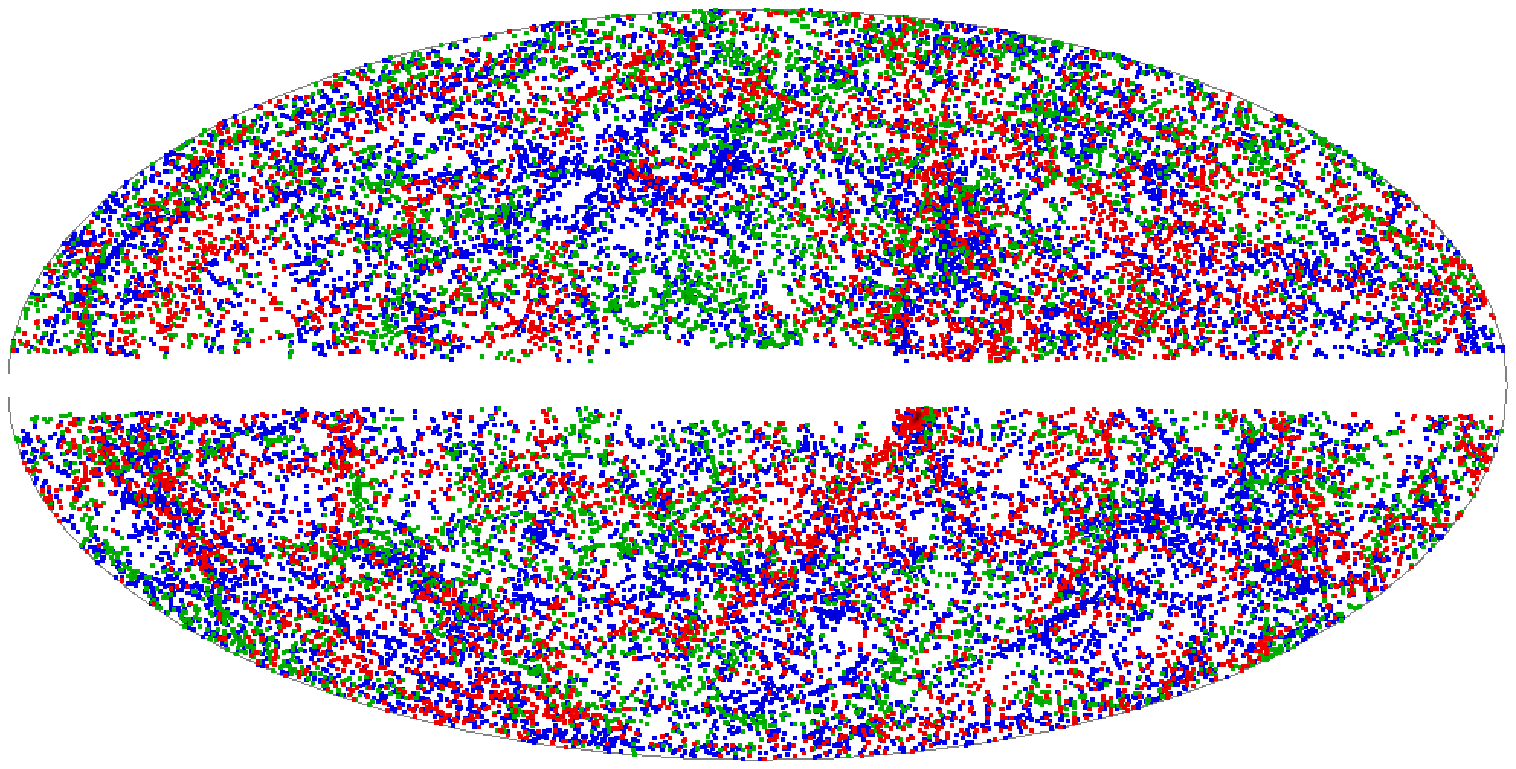}
\includegraphics[height=7.0cm,width=14cm,angle=0]{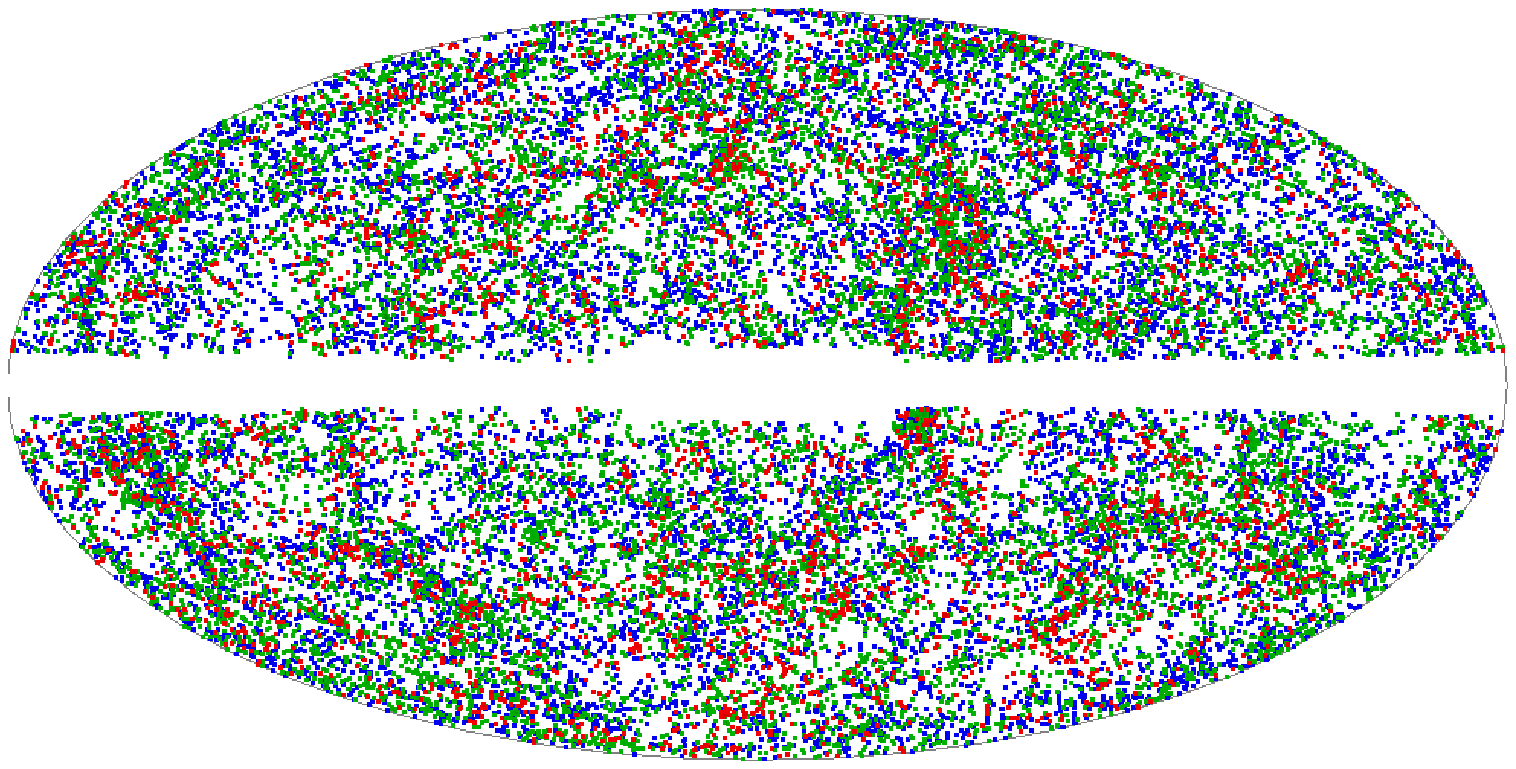}
\caption{The distribution of 2MRS groups in Galactic coordinates, with
  Galactic longitude increasing from $0^\circ$ at the center to
  $180^\circ$ to the left, and from $180^\circ$ from the right to
  $360^\circ$ at the center.  The Galactic latitude from $-90^\circ$
  to $90^\circ$ from bottom to top.  Upper panel: red, green and blue
  points represent groups within the redshift range: $0.0\leq z
  \leq0.02$, $0.02\leq z \leq0.03$ and $0.03\leq z \leq0.08$,
  respectively. Lower panel: red, green and blue points represent
  groups within the mass range:$\log M_{\rm 2MRS}\ge 13.5$, $13.5 \ge
  \log M_{\rm 2MRS} \ge 12.5$ and $12.5 \ge \log M_{\rm 2MRS}$,
  respectively. }
\label{fig:distribution}
\end{figure*}

\begin{figure*}
\center
\vspace{0.5cm}
\includegraphics[height=6.0cm,width=15.0cm,angle=0]{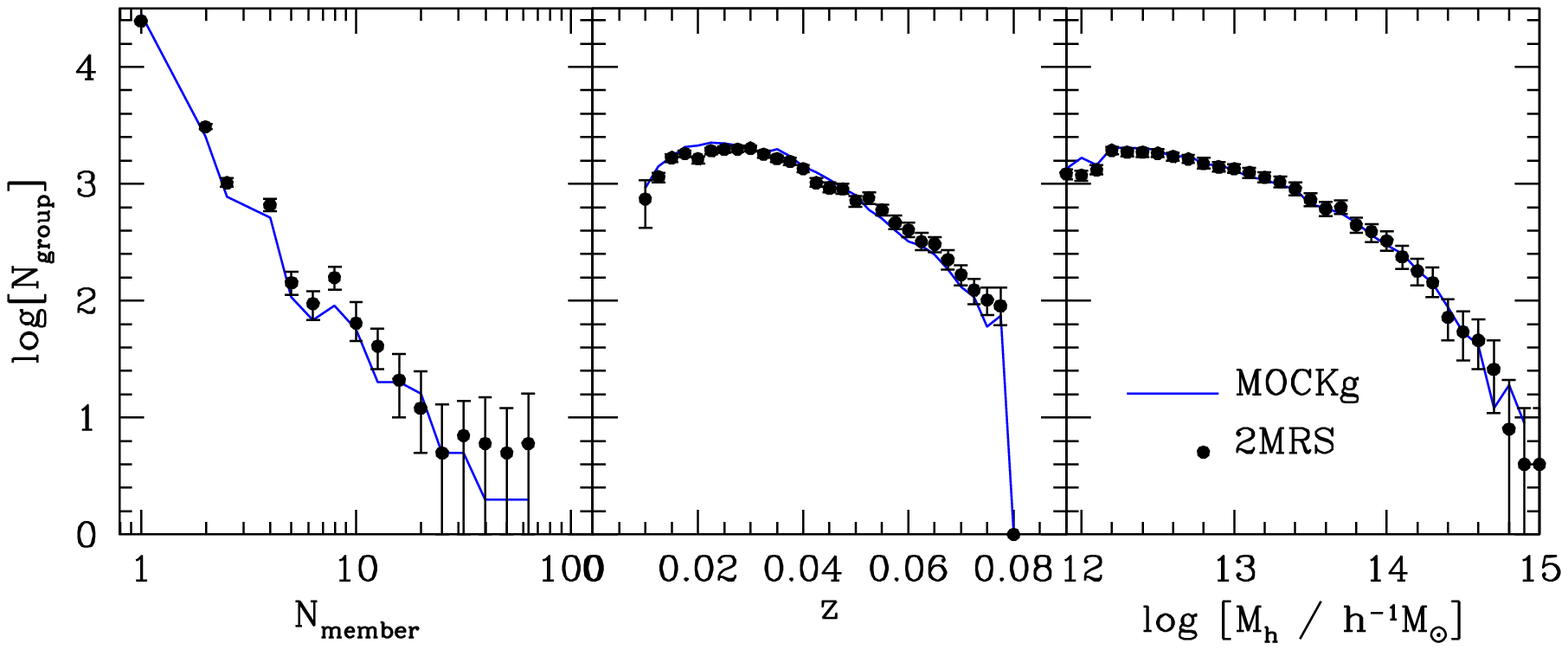}
\caption{The number of groups as function of the number of group
  members (left panel) ,group redshift (middle panel) and halo mass
  (right panel).  The solid points with error bars show the results
  obtained from the 2MRS group catalog constructed using the modified
  halo-based group finder. The error bars are given by 1000 bootstrap
  re-samplings. For comparison, in all panels, we also plot the
  corresponding distributions obtained from the mock 2MRS group
  catalog (MOCKg) using curves. }
\label{fig:distri}
\end{figure*}

We apply our modified group finder to the 2MRS galaxy catalog in
exactly the same way with MOCKg catalog as described in the last
section.  In the following, we describe our catalog and present some
of its basic properties.  We also make comparisons with the SDSS
groups in the overlapping region, and discuss how some known nearby
structures are represented in our catalog.

\subsection{Basic Properties}

\begin{deluxetable*}{lcccccc}
 \tabletypesize{\scriptsize} \tablecaption{Properties of
   2MRS catalogs } \tablewidth{0pt}

  \tablehead{Sample & Galaxies  & Groups   & $N = 1$   &
    $N \ge 2$   & $14.0 \ge \log M_h \ge 13.0$ & $\log M_h\ge 14.0$} \\
  \startdata
  2MRS $0.01\le z \le 0.03$ & $20921$ &  $12879$ &  $10004$ &  $2875$ &  $1495$  & $61$  \\
  2MRS $0.0 \le z \le 0.08$ & $43246$ &  $29904$ &  $24618$ &  $5286$ &  $8484$ & $1103$   \\
  MOCKg    & $41876$ &  $32368$ &  $28143$ &  $4225$ &  $8085$ & $1098$ \\
  MOCKt    & $41876$ &  $34846$ &  $31879$ &  $2967$ &  $6103$ &  $867$
 \enddata
\label{tab:properties}
\end{deluxetable*}

Our modified halo-based group finder identifies 29,904 groups from a
total of 43,246 2MRS galaxies in the redshift range $z \leq
0.08$. Among the groups selected, 5,286 have two or more members;
2,208 are triplets; and 1,189 have four or more members.
Fig. \ref{fig:distribution} shows the distribution of all groups in
the 2MRS catalog.  In the upper panel, the red points represent groups
in the redshift range $0.0 < z \leq 0.02$, while green and blue points
represent groups in $0.02 < z \leq 0.03$ and $0.03 < z \leq 0.08$,
respectively.  In the lower panel, red, green and blue points
represent groups in mass ranges $\log M_{\rm 2MRS}\ge 13.5$, $13.5 \ge
\log M_{\rm 2MRS} \ge 12.5$ and $12.5 \ge \log M_{\rm 2MRS}$,
respectively. One can see from the lower panel that more massive
groups seem to locate preferentially denser regions.

Table \ref{tab:properties} lists the number of groups in the 2MRS
within two redshift ranges: $z = 0.01 - 0.03$ and $z \leq 0.08$, with
single member or with more than one member. We also list the number of
massive groups with estimated halo masses in two mass ranges, $14.0
\ge \log M_{\rm 2MRS} \ge 13.0$ and $\log M_{\rm 2MRS} \ge 14.0$.  For
comparison, the number of the mock groups constructed by our
  group finder (MOCKg) and given by simulation (MOCKt) in the
redshift range $z \leq 0.08$ are also listed in Table
\ref{tab:properties}.  We show in the left panel of Fig.
\ref{fig:distri} the richness distribution of groups in 2MRS which are
shown as the dots with error bards.  Compare to the MOCKg which is
shown as the solid line, the 2MRS sample tends to contain more rich
groups with $N_g>32$. However, the total number of such rich systems
is small and the statistic is rather poor. The middle panel of Fig.
\ref{fig:distri} shows the redshift distribution of groups, where the
redshift of each group is the luminosity-weighted average of the
redshifts of its member galaxies. Here we see that the 2MRS sample
contains slightly less groups at low redshift and slightly more groups
at high redshift than the mock 2MRS sample.

One of the purposes of constructing the 2MRS galaxy group catalog is
to populate the local Universe with well estimated dark matter halos
for our subsequent reconstructions of the local density field
\cite[e.g. ][]{Wang2014}.  We check the number distribution of galaxy
groups as a function of halo mass in the 2MRS volume which are shown
in the right panel of Fig. \ref{fig:distri} using black points with
error bars.  For comparison, the ones obtained form the MOCKg
groups are also plotted in this figure. The good agreement
  between MOCKg and 2MRS indicate that the number distribution of
  groups as a function of halo mass beyond the redshift completeness
  $z_{\rm limit}$ are also quite similar in the MOCKg and 2MRS
  samples.

\subsection{Comparisons with previous results}

\begin{figure*}
\center
\vspace{0.5cm}
\includegraphics[height=12.0cm,width=13.5cm,angle=0]{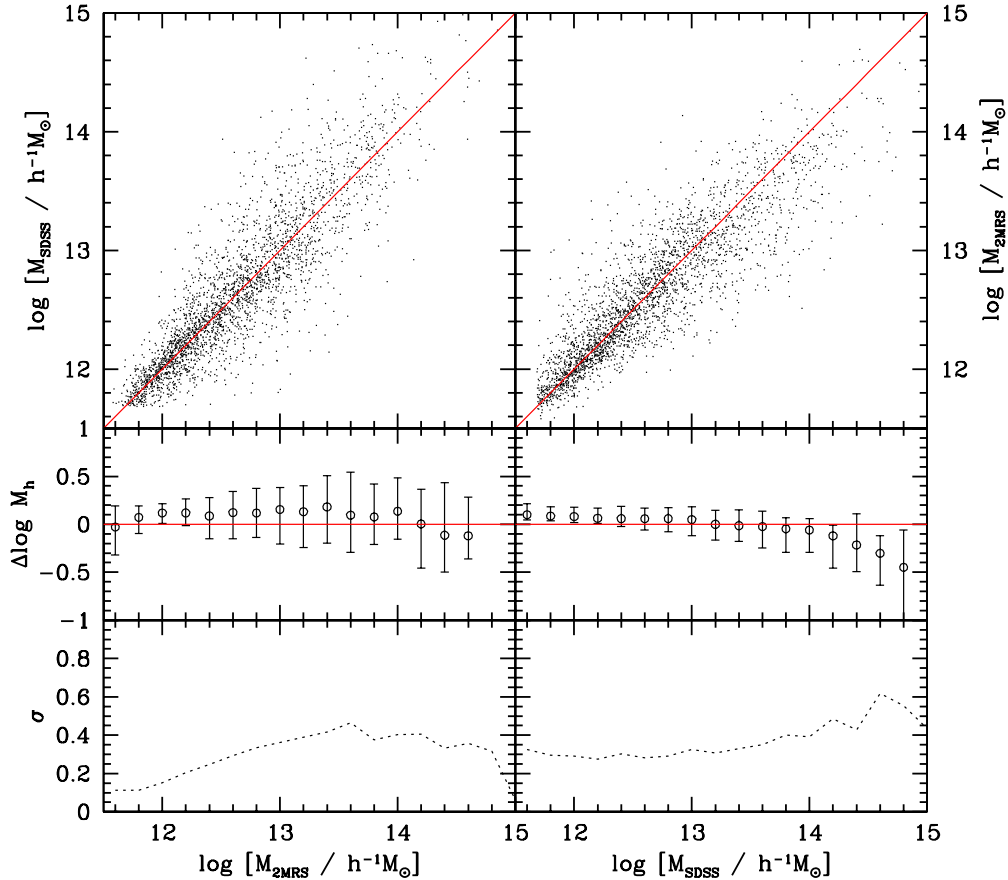}
\caption{Comparison of the estimated halo mass between 2MRS and SDSS
  DR7 galaxy groups.  This is similar to Fig. \ref{fig:Mhtestmock},
  but with the true halo masses replaced by SDSS DR7 halo masses. The
  difference between the two halo masses, $\Delta \log M_h = \log
  M_{\rm 2MRS} - \log M_{\rm SDSS}$ are also shown in the middle row
  in bins of halo mass obtained from the 2MRS (left) and SDSS (right),
  respectively.  The error bars indicate the $16\%-84\%$ percentiles
  of the distributions. The standard variation of groups from the
  redlines in the top two panels are shown as the $\sigma$ curves
  plotted in the bottom panels. }
\label{fig:Mcompare}
\end{figure*}

\begin{deluxetable}{lccc}
 \tabletypesize{\scriptsize} \tablecaption{Comparison between
 halo masses (in $\log [M_h/\msunh]$)} \tablewidth{0pt}

  \tablehead{Groups &  Lu  & T15 & Literature }  \\
  \startdata
  Abell2199   & 14.84 & 15.26  & 14.81$^{1}$ \\
  Coma        & 14.76 & 15.23  & 14.85$^{2}$ \\
  Abell2634   & 14.59 & 14.90  & 14.61$^{3}$ \\
  Perseus     & 14.59 & 15.07  \\
  Norma       & 14.30 & 15.10  & 15.00$^{4}$ \\
  Virgo       & 14.37 & 15.04  & 14.43-14.90$^{5}$ \\
  Abell1367   & 14.34 & 14.80
 \enddata
 \tablecomments{$^{1}$ \citet{Kubo2009}. $^{2}$ \citet{Gavazzi2009}.
   $^{3}$ \citet{Schindler1997}. $^{4}$ \citet{Woudt2008}. $^{5}$
   \citet{Karachentsev2010}.  }
\label{tab:mhcompare}
\end{deluxetable}

In a recent study, \citet[][hereafter T15]{Tully2015} identified
galaxy groups from the 2MRS using a modified version of the halo-based
group finder developed by \citet{Yang2005a}, with halo masses
estimated from a scaling relation to the characteristic group
luminosity.  Tully identified 13,606 groups from a total 24,044
galaxies in the velocity range $3000 - 10,000 \kms$, among which 3,461
have more than one member.  In comparison, our group catalog uses a
different halo mass estimator and extends to a larger redshift range.
In particular, we have used a realistic mock catalog to quantify the
reliability of our group finder and the group masses it gives.

To compare with T15, we list in Table. \ref{tab:properties} the
properties of groups in the redshift range $0.01 \leq z\leq 0.03$,
which is comparable to the redshift range used in T15.  For a total of
20,921 galaxies, we identified 12,889 groups, which matches well with
the results of T15. The richest group has 184 member galaxies in our
catalog, which is consistent with 180 member galaxies given by T15.
We list the estimated halo masses of some prominent nearby groups in
Table. \ref{tab:mhcompare}, including groups in the
  Perseus-Pisces filament, Leo cluster, Norma cluster, Virgo and Coma
  clusters. For comparison, the results given by T15 for the same
  groups are also listed in the table. In general, the halo masses
given by T15 tend to be larger than the masses we obtain. We suspect
that this is caused by different definitions of halo masses.  To
investigate this further, we looked into the literature for the halo
masses of the groups in question, and the results are also shown in
Table. \ref{tab:mhcompare}.  In general our mass estimates match well
with the values given in the literature. The only exception is the
Norma cluster, for which our mass estimate is significantly
lower. However, Norma is located near the Milky Way Zone of Avoidance,
and is severely obscured by the interstellar dust at the optical
wavelengths. It is unclear if this is also a significant problem in
the near infrared data used here.

We further test our 2MRS group catalogs by comparing with an existing
group catalog.  This group catalog used here was constructed by Y07
from the New York University Value-Added Galaxy Catalog
\citep[NYU-VAGC; ]{Blanton2005b} based on the SDSS Data Release 7.  A
total of 639,359 galaxies with redshifts $0.01 \leq z \leq 0.20$ and
redshift completenesses $C > 0.7$ were selected for constructing their
group catalog. They found a total of 472,416 groups, among which
23,700 have three or more members.  For our comparison, we first cross
match the 2MRS galaxies with te SDSS DR7 galaxies according to their
coordinates in the sky.  With the assumption that galaxies located
within $5 \arcsec$ of one another in the sky, and with a redshift
difference of $\Delta z < 5\times10^{-4}$ (corresponding to a velocity
of 150 km/s) are the same one, we got a total of 4,528 galaxy pairs,
among which 2,938 galaxy pairs are centrals in both group catalogs.

We investigate the estimated halo masses assigned to the same halo in
the two galaxy group catalogs. Here halos from the two group catalogs
are matched if they have the same central galaxy according to the
matched galaxy pairs. Note that, for the SDSS galaxy groups, the halo
masses are estimated by the `RANK' method, which estimates the halo
mass of a candidate galaxy group according to its characteristic
luminosity, $L_{-19.5}$, defined as the total luminosity of member
galaxies brighter than a given luminosity threshold $\rmag = -19.5$.
The top panels in Fig. \ref{fig:Mcompare} show the estimated halo
masses for all the 2,938 matched central galaxy pairs given by the
2MRS and SDSS DR7 group catalogs, respectively.  The top left and top
right panels plot the same thing, except that the two mass axes are
flipped: SDSS mass versus 2MRS mass in the left and 2MRS mass versus
SDSS mass in the right). Ideally, the two estimated halo masses should
be the same, so that all the data points would lie on the red solid
line ($\log M_{\rm 2MRS} = \log M_{\rm SDSS}$).  We can see that the
two halo masses estimated are tightly correlated with each another,
with no obvious systematic bias (see the middle two panels which show
the deviations from the perfect line).  The typical scatter is $\sim
0.4{\rm dex}$ in medium to massive halo mass range, and $\sim 0.2$ -
$0.3 {\rm dex}$ for low mass groups, as shown in the two lower panels.
This scatter is roughly consistent with the one shown in
Fig. \ref{fig:Mhtestmock} between the group and true halo masses
estimated from the mock 2MRS catalogs.

\section[]{SUMMARY}
\label{sec:summary}

In this paper, we have implemented, tested and applied a modified
version of the halo-based group finder developed in \citet{Yang2005a,
  Yang2007} to extract galaxy groups from the 2MRS.
Covering uniformly about 91\% of the sky,  the 2MRS provides the best available
representation of the structures in local universe, and so a group catalog constructed
from it is useful for many purposes. However, 2MRS is quite shallow; in
many cases only a few brightest members within a halo can be observed.
To deal with this limit, we have updated the halo mass estimate
used in the previous group finder with a new method based on
`GAP'.  This `GAP' estimator consists of two parts: (i) a relation between
the luminosity of the central galaxy, $L_c$,  and the halo mass, $M_h$, inferred
iteratively from abundance matching between
the luminosity of {\it central} galaxies and the masses of dark matter
halos; (2) a luminosity gap correction factor obtained from the
luminosity difference between the central galaxy and a faint
satellite galaxy.

In order to evaluate the performance of our modified group finder, we
have constructed mock 2MRS galaxy samples based on the
observed $K_s$-band luminosity function.  The group catalog obtained from
the mock 2MRS galaxy catalog shows a 100\% completeness
for about $65\%$ of the most massive groups to $\sim85\%$
for groups with halo masses $\log M_h < 10^{14}\msunh$.
On average, about 80\% of the groups have 80\% completeness.  In terms of
interlopers, about $65\%$ of the groups identified have none, and an additional
20\% have an interloper fraction lower than 50\%.
Further tests on the halo mass estimation show that the
deviation of the halo mass between the selected groups and the true
halos is $\sim 0.35 {\rm dex}$ over the entire mass range.
These tests demonstrate that the modified group finder is reliable
for the 2MRS sample.

 Applying the modified halo-based group finder to the 2MRS, we have
obtained a group catalog with a depth to  $z \leq 0.08$ and covering
$91\%$ of the whole sky. This 2MRS group catalog contains a total
of 29,904 groups, among which 24,618 are singles and 5,286
have more than one member. Some of the basic properties of the
group catalog are presented, including the distributions in richness,
in redshift and in halo mass.  This catalog provides a useful data base to
study galaxies in different environments.  In particular,  it can be used to
reconstruct the mass distribution in the local Universe, as we will do
in a forthcoming paper.

\section*{Acknowledgements}

We thank the anonymous referee for helpful comments that greatly
improved the presentation of this paper.  This work is supported by
973 Program (No. 2015CB857002), national science foundation of China
(grants Nos. 11203054, 11128306, 11121062, 11233005, 11073017),
NCET-11-0879, the Strategic Priority Research Program ``The Emergence
of Cosmological Structures" of the Chinese Academy of Sciences, Grant
No. XDB09000000 and the Shanghai Committee of Science and Technology,
China (grant No. 12ZR1452800).  We also thank the support of a key
laboratory grant from the Office of Science and Technology, Shanghai
Municipal Government (No. 11DZ2260700).  HJM would like to acknowledge
the support of NSF AST-1517528.

A computing facility award on the PI cluster at Shanghai Jiao Tong
University is acknowledged. This work is also supported by the High
Performance Computing Resource in the Core Facility for Advanced
Research Computing at Shanghai Astronomical Observatory.

\label{lastpage}

\end{document}